\documentclass[aps,prd,preprint,groupedaddress,showpacs]{revtex4-1}

\usepackage{graphicx}						
\usepackage{amsmath,amsfonts,amsbsy}
\usepackage{slashed} 

\declareslashed{}{/}{.15}{0}{\laser}

\newcommand{\laser}{\mathcal{A}}
\newcommand{\Ac}{{A^\mathrm{c}}}
\newcommand{\ee}{\mathrm{e}}
\newcommand{\ecc}{\tau}   
\newcommand{\eccp}{\ecc^{+}\!}
\newcommand{\eccm}{\ecc^{-}\!}
\newcommand{\psl}{\slashed{p}}

\newcommand{\dsl}{\slashed{\partial}}
\newcommand{\Asl}{\slashed{\Ac}}
\newcommand{\ksl}{\slashed{k}}
\newcommand{\msl}{\!\slashed{\,m}}
\newcommand{\mslb}{\slashed{\bar{{m}}}}

\newcommand{\Vsl}{\slashed{\laser}}
\newcommand{\Vssl}{\slashed{\laser}^{*}}
\newcommand{\J}{J^{\ecc}}
\newcommand{\Js}{J^{\ecc *}}

\newcommand{\pbar}{{\bar{p}}}
\newcommand{\qbar}{{\bar{q}}}
\newcommand{\pbsl}{\slashed{\pbar}}

\newcommand{\U}{\mathcal{\,U}}
\newcommand{\V}{\mathcal{V}}
\newcommand{\Uv}{\mathcal{\,U}_{_{\mathrm{V}}}}

\newcommand{\Vv}{\mathcal{V}_{_{\mathrm{V}}}}
\newcommand{\Uvb}{\bar{\mathcal{\,U}}_{_{\mathrm{V}}}}

\newcommand{\Vvb}{\bar{\mathcal{V}}_{_{\mathrm{V}}}}
\newcommand{\F}{\mathcal{F}}

\newcommand{\psiV}{\psi_{_{\mathrm{V}}}}
\newcommand{\psibV}{\bar{\psi}_{_{\mathrm{V}}}}

\newcommand{\aVa}{a^{(\alpha)}_{_{\mathrm{V}}}\!}

\newcommand{\bVb}{b^{(\beta)}_{_{\mathrm{V}}}\!}

\newcommand{\adVb}{a^{\dag(\beta)}_{_{\mathrm{V}}}\!}
\newcommand{\bdVa}{b^{\dag(\alpha)}_{_{\mathrm{V}}}\!}

\newcommand{\kdotx}{k{\cdot}x}

\newcommand{\De}{D}

\newcommand{\intfps}{\int{\bar{}\kern-0.45em d}^{\,3}p\,}
\newcommand{\intps}{\int\frac{{\bar{}\kern-0.45em d}^{\,3}p\,}{2E^*_p}}
\newcommand{\intfp}{\int{\bar{}\kern-0.45em d}^{\,4}p\,}
\newcommand{\intfqs}{\int{\bar{}\kern-0.45em d}^{\,3}q\,}
\newcommand{\intfpqs}{\int{\bar{}\kern-0.45em d}^{\,3}p\,\,\,{\bar{}\kern-0.45em d}^{\,3}q\,}

\newcommand{\ket}[1]{|#1\rangle}
\newcommand{\braV}[1]{{}_{_{\mathrm{V}}}\!\langle #1|}
\newcommand{\ketV}[1]{|#1\rangle_{_{\mathrm{V}}}}

\newcommand{\kx}{k{\cdot}x}
 \newcommand{\ky}{k{\cdot}y}
\begin{document}

\title{Electrons in an eccentric background field}

\author{Martin~Lavelle and David~McMullan}
\affiliation{Centre for Mathematical Sciences\\Plymouth  University\\
Plymouth, PL4 8AA, UK}

\date{\today}

\begin{abstract}
We present a description of  electrons propagating in an elliptically polarised, plane wave background which includes circular and linear polarisations as special cases. We calculate, to all orders in the background field, the two point function and relate it to various expressions found in the literature. The background field induced mass shift of the electron is shown to be polarisation independent in the full elliptic class. The matrix nature of this mass shift in the fermionic theory is discussed. The extent to which a momentum space description is possible for this system is clarified.

\end{abstract}



\maketitle

\section{Introduction}

High intensity lasers are becoming an attractive testing ground for exploring new regimes in the standard model~\cite{Heinzl:2011ur}\cite{DiPiazza:2011tq}. The Volkov description~\cite{Volkov:1935zz} of an electron in a  plane wave background provides the equivalent of the free solution in quantum electrodynamics. It is an all orders, exact solution displaying the key new features for this type of intense interaction. They include a laser induced electron mass shift and the loss of translational invariance~\cite{Reiss:1966A}\cite{Brown:1964zz}\cite{Neville:1971uc}\cite{Dittrich:1973rn}\cite{Dittrich:1973rm}\cite{Kibble:1975vz}\cite{Mitter:1974yg}\cite{Ritus:1985review}\cite{Ilderton:2012qe}\cite{Lavelle:2013wx}\cite{Lavelle:2014mka}\cite{Lavelle:2015jxa}.  Despite recent progress in modelling more realistic laser pulses~\cite{Fedotov2009}\cite{DiPiazza:2013vra}\cite{DiPiazza:2015xva}\cite{DiPiazza:2016tdf}\cite{Heinzl:2017zsr}\cite{Waters:2017tgl}\cite{Karbstein:2017jgh}, the plane wave solution for matter fields in a laser is still widely used and provides much insight into the physics of matter in extreme conditions.  

The Volkov solution has been studied extensively for circular and linear laser polarisations in both scalar and fermionic versions of electrodynamics. These polarisations can be understood as limiting cases of the wider class of elliptical polarisations. We note that vacuum birefringence, through polarisation flipping \cite{king2016measuring} \cite{king2016vacuum}, is supposed to generate the  wider elliptical class from, say, an initial linear polarisation. 

 In this paper we will study electrons in an elliptically polarised plane wave background and through this will identify which structures in the Volkov solution depend upon 
the eccentricity of the polarisation. We will also see that the total averaged electromagnetic energy is independent of the eccentricity.

In section~\ref{sec:field} we will introduce the elliptically polarised laser field and relate our description in terms of the eccentricity to the Stokes' parameters. Then, in section~\ref{sec:quantised}, we will solve the Dirac equation in a general elliptical background and show that the mass shift is independent of the eccentricity of the polarisation. We will consider linear and circular polarisations as limits of our general case. Following this, in section~\ref{sec:twopoint}, we will calculate the two point function for the electron in this general background and show how to make contact between various formulations in the literature. Two appendices contain details of the calculations and a new class of Bessel functions which are required to describe the full elliptical class of polarisations.  

\newpage

\section{The Elliptic Laser Field}\label{sec:field}
An elliptically polarised plane wave with null momentum $k_\mu$ is described by the classical  potential $\Ac$ where
\begin{equation}\label{eq:Aa}
  \Ac^\mu(x)=a_1^\mu\ecc^{{+}}\!\cos(\kx)+a_2^\mu\ecc^{{-}}\!\sin(\kx)\,.
\end{equation}
The orthogonal amplitudes in this potential are taken to satisfy the light cone gauge conditions that
\begin{equation}\label{eq:lcgauge}
  k\cdot a_1=k\cdot a_2=0
\end{equation}
and the spacelike normalisation that 
\begin{equation}\label{eq:normalisation}
  a_1\cdot a_1=a_2\cdot a_2=a^2<0\,.
\end{equation}
The eccentricity parameters in~(\ref{eq:Aa}) are defined by $\ecc^{\pm}=\sqrt{1\pm\ecc^2}$, where $\ecc$ determines the overall eccentricity of the laser field. In particular,   circular polarisation corresponds to $\ecc=0$ and linear to $\ecc=1$. 
    
The  geometric eccentricity of an ellipse is related to the above eccentricity parameter  by identifying the major axis as pointing along the direction of the vector $a_1^\mu$, the minor one is then along the vector  $a_2^\mu$. Using standard properties of such  ellipses, we have
\begin{align}\label{eq:geo_ecc}
  \mathrm{geometric\ eccentricity}&=\sqrt{
  \frac{
  a_1{\cdot}a_1(\ecc^{{+}})^2-a_2{\cdot}a_2(\ecc^{{-}})^2
       }{a_1{\cdot}a_1(\ecc^{{+}})^2}} 
=\sqrt{\frac{2\ecc^2}{1+\ecc^2}}\,.
\end{align}
From this we see that the eccentricity parameter $\ecc$ is less than the geometric eccentricity for all polarisations between the circular and linear limits.

One of the key physical consequences of this description of the full elliptic class of polarisations is that the total averaged electromagnetic energy is independent of the eccentricity. This follows directly from the electromagnetic potential (\ref{eq:Aa}) since the electric and magnetic fields can be easily calculated and squared. Averaging over a laser cycle will then remove eccentricity dependent cross terms and result in an expression proportional to 
$(\ecc^{{+}})^2+(\ecc^{{-}})^2$, which is independent of $\ecc$. 

The single, unified framework of~(\ref{eq:Aa}) describes an experimental set up with fixed laser energy for any polarisation in the elliptic class. Note that, within this approach, the averaged value of the field amplitude,  $(\Ac)^2$, is polarisation dependent. If we had kept  the averaged amplitude fixed within this class, then the averaged energy would be polarisation dependent. 

The expression (\ref{eq:Aa}) for the background laser field, given in terms of the  amplitudes $a_1$, $a_2$ and the  eccentricity parameters, can be  simplified by introducing the  complex laser amplitude  $\laser ^\mu$ defined by
\begin{align}\label{eq:Vdef}
\begin{split}
  e\Ac^\mu(x)=&\tfrac12e(a_1^\mu\ecc^{+}+i a_2^\mu\ecc^{-})\ee^{-i\kx}+\tfrac12e(a_1^\mu\ecc^{+}-i a_2^\mu\ecc^{-})\ee^{i\kx}\\
  :=& \laser^{\mu}\ee^{-i\kx}+\laser^{*\mu}\ee^{i\kx}\,. 
\end{split}
\end{align}
Here we  have absorbed the coupling into the complex laser amplitude  to make clear that the relevant expansion parameter is proportional to $ea$, and that the size of this determines whether we are in the weak field, $|\laser|<m$, or strong field, $|\laser \vert>m$, regime where $m$ is the electron mass.
  
In terms of  this complex amplitude the  original real amplitudes and eccentricity parameters are recovered by noting that $ea_1^\mu\eccp=\laser ^{*\mu}+\laser^\mu$ and $ea_2^\mu\eccm=i(\laser^{*\mu}-\laser^\mu)$. The modulus of this amplitude is particularly simple since
\begin{equation}\label{eq:vvstar}
   2\laser{\cdot}\laser^*=e^2a^2\,,
\end{equation}
which we note is independent of the polarisation. This will be important below when we look at the mass of the associated charged particles. We also note that in the limit of linear polarisation $\laser^\mu$ is real while for all other polarisations it is complex. In addition, due to the orthogonality of the real amplitudes in the laser potential, $\laser^2$ is real for all polarisations and vanishes for circular polarisation. 

This description of the laser field directly in terms of the eccentricity parameter will allow for a direct route to the quantum theory, but this is not the  way such laser configurations are usually discussed, see for example~\cite{berestetskii2012quantum}, \cite{mcmasters1954pol} and \cite{collett2005field}. To translate into the more familiar  Stokes' parameter approach to polarisation, we recall that the amplitudes in the potential~(\ref{eq:Aa}) satisfy the orthogonality condition (\ref{eq:lcgauge}). If we take the laser to be pointing in the $x^3$ direction, then we can write
\begin{equation}\label{eq:a1a2}
    a_1=\begin{pmatrix}
      0\\\alpha\\\beta\\0
    \end{pmatrix}+\Lambda_1 \begin{pmatrix}
      1\\0\\0\\1
    \end{pmatrix}\qquad\mathrm{and}\qquad
 a_2=\pm\begin{pmatrix}
      0\\\beta\\-\alpha\\0
    \end{pmatrix}+\Lambda_2 \begin{pmatrix}
      1\\0\\0\\1
    \end{pmatrix}\qquad
  \end{equation}  
where $\Lambda_1$ and $\Lambda_2$ are undetermined constants. From (\ref{eq:normalisation}) we see that  $\alpha^2+\beta^2=-a^2$. 

Inserting these expressions into the description of the complex laser amplitude~(\ref{eq:Vdef}),  allows us to identify the polarisation density tensor for this elliptically polarised laser as
\begin{align} \label{eq:stokes}
  \rho
  :=-\frac1{\laser{\cdot}\laser^*}\begin{pmatrix}
    \laser^1\laser^{1*}&\laser^1\laser^{2*}\\\laser^2\laser^{1*}&\laser^2\laser^{2*}
  \end{pmatrix}=\frac1{2I}\begin{pmatrix}
    I+Q&U-iV\\U+iV&I-Q
  \end{pmatrix}
\end{align}
where $I=\alpha^2+\beta^2$, $Q=\ecc^2(\alpha^2-\beta^2)$, $U=\ecc^22 \alpha\beta $ and $V=\mp\sqrt{1-\ecc^4}I$. These parameters, here generalised for the full elliptic class, are called the Stokes' parameters.

This polarisation  tensor  is not invariant under rotations in the transverse plane and under such a transformation through an angle $\theta$, represented by the matrix  $R(\theta)$, we have $\rho\to\rho'= R^\dag(\theta)\rho R(\theta)$ where
\begin{equation}
  \rho'=\frac1{2I}\begin{pmatrix}
    I+Q'&U'-iV\\U'+iV&I-Q'
  \end{pmatrix}
\end{equation}
with $Q'=Q\cos2\theta+U\sin2\theta$ and $U'=U\cos2\theta-Q\sin2\theta$. This rotation of two of the Stokes' parameters and the invariance of the third, is best understood by looking at the circular and linear limits.

For circular polarisation, where $\ecc=0$, we have $Q=U=0$ and $V=\pm I$. The choice in sign of $V$ corresponds to either left ($V=I$) or right ($V=-I$) circular polarisation.  Clearly, a rotation will not affect this result, and this is encoded in the above invariance of $V$.

In the linear limit, where $\ecc=1$, we have $V=0$ while 
\begin{equation}
  Q=\frac{\alpha^2-\beta^2}{\alpha^2+\beta^2}\,,\qquad U=\frac{2\alpha\beta}{\alpha^2+\beta^2}\,.
\end{equation}
Now through a rotation we can set $\beta=0$, in which case  $Q=I$ and $U=0$ giving a horizontal polarisation.  Or we can rotate so that $\alpha=0$, which does not change $U=0$ but flips $Q$ so that  $Q=-1$, corresponding to a vertical polarisation. Alternatively, we can rotate so that $\alpha=\pm\beta$ so that now $Q=0$ while $U=\pm I$ and this corresponds to a linear polarisation of $\pm45^{\circ}$. 

Our expression (\ref{eq:stokes}) for the polarisation tensor has thus allowed for a precise identification of the Stokes' parameters for this full elliptic class. We have also seen how to recover the more familiar circular and linear eccentricity limits for these parameters.  We shall see in the next section that quantising in this background is best addressed using the eccentricity parametrisation of the polarisation rather than the Stokes' description. The dictionary implicit in (\ref{eq:stokes}) allows a simple translation back into the parameters more familiar to the wider laser community.

\section{Quantised Matter in a Laser}\label{sec:quantised}

The  interaction of  matter with the plane wave background (\ref{eq:Aa}) is achieved through the usual minimal coupling prescription whereby the derivative $\partial_\mu$ is replaced by the covariant derivative $\partial_\mu+ie \Ac_{\!\mu}$ in the equations of motion for the matter. The structure of the potential (\ref{eq:Aa}) allows for an exact description of this interacting theory~\cite{Volkov:1935zz}, and we will refer to this as the Volkov solution. The route to the quantum description of such an interacting theory  will now be based upon this Volkov solution rather than the free theory.

In references \cite{Lavelle:2013wx} and \cite{Lavelle:2015jxa} the quantisation of both scalar and fermionic matter propagating in circular and linearly polarised backgrounds were addressed, but the analysis was restricted to the translationally invariant part of the two point function. We now extend that analysis to both the full elliptic class and the complete  two point function. Through this we shall identify the eccentricity dependencies of both the  mass and wave function normalisation, and delineate the impact of the laser background on translational invariance.

In this paper we consider only the interactions of the electron with the background field, i.e., we neglect all photons that are not degenerate with the laser background. Working in the Heisenberg picture appropriate to these interactions, the Volkov field  $\psiV(x)$   satisfies the coupled Dirac equation
\begin{equation}\label{eq:eqm}
  i(\dsl+ie\Asl)\psiV(x)=m\psiV(x)\,,
\end{equation} 
where $\dsl$, for example, is the Feynman slash notation, $\dsl=\partial_\mu \gamma^\mu$. 
Recall that in the free theory, the Dirac equation is solved by Fourier expanding in the on-shell momentum, $p_{_{\mathrm{OS}}}^\mu$, where $p^2_{_{\mathrm{OS}}}=m^2$. The operator content of the field is then carried by the creation and annihilation operators, $a^{(\alpha)}(p)$ and $b^{\dag{(\alpha)}}(p)$, multiplied by appropriate spinorial terms, $\U^{(\alpha)}(p)$ and $\V ^{(\alpha)}(p)$. In the Volkov background, the solution to~(\ref{eq:eqm}) has a similar structure 
and we find
\begin{align}\label{eq:fvolkov}
\begin{split}
  \psiV(x)=\intfps\frac{m}{E^*_p}&\Big(\F(x,\pbar)\De(x,\pbar)\Uv^{(\alpha)}(p)\aVa(p)\\&\qquad\qquad+\F(x,-\pbar)\De(x,-\pbar)\Vv^{(\alpha)}(p)\bdVa(p)\Big)\,,
\end{split} 
\end{align}
where the integration measure notation absorbs factors of $2\pi$ so that $\kern0.45em{{\bar{}\kern-0.45em d}^{\,3}p}=d^3p/(2\pi)^3$.
The adjoint field is 
\begin{align}\label{eq:fbarvolkov}
\begin{split}
  \psibV(y)=\intfqs\frac{m}{E^*_q}&\Big(\Uvb^{(\beta)}(q)\F(y,-\qbar)\De(y,\qbar)\adVb(q)\\
  &\qquad\qquad +\Vvb^{(\beta)}(q)\F(y,\qbar)\De(y,-\qbar)\bVb(q)\Big)\,.
\end{split}
\end{align} 
We will now unpick the details of this solution. The Volkov spinors $\Uv$ and $\Vv$ are defined in  Appendix~\ref{sec:appen1} along with the Volkov creation and annihilation operators. 

Before looking at the fermionic pre-factor, $\F(x,\pbar)$, and distorted plane wave term, $\De(x,\pbar)$, in~(\ref{eq:fvolkov}), we note that in these expressions the momentum $\pbar=(E^*_p,\underline{p})$ satisfies the shifted on-shell condition of the so called quasi-momentum:
\begin{equation}\label{eq:on_shell}
  \pbar^2-m^2_*=E_p^{*\,2}-\underline{p}^{\,2}-m^2_*=0\,,
\end{equation}
where  
\begin{equation}\label{eq:mass_star_square}
  m^2_*=m^2+\delta m^2=m^2-2\laser{\cdot}\laser ^*\,.
\end{equation}
An important feature, revealed here by studying the full elliptic class, is that the mass shift seen in~(\ref{eq:mass_star_square})  is  polarisation independent due to~(\ref{eq:vvstar}) so that, for the full elliptic class,
\begin{equation}
  \delta m^2=-e^2 a^2\,.
\end{equation}

 This  result is not evident when the extreme cases of  circular and linearly polarised lasers are  considered separately. A  mass-shift of $\delta m^2=-\frac12e^2a^2$ is commonly quoted for linearly polarised lasers, while for circularly polarised lasers a mass shift of $\delta m^2=-e^2a^2$ is given. This apparent difference can be  traced back to different normalisations of the two laser fields and hence  the total energy being considered. Within the elliptic class, using a common total energy, the mass shift is seen to be independent of the eccentricity. 

The fermionic pre-factor $\F(x,\pbar)$ in (\ref{eq:fvolkov})  is found to be
\begin{equation}\label{eq:Fdefinition}
  \F(x,\pbar)=1+e\frac{\ksl\Asl(x)}{2\pbar{\cdot} k}\,.
\end{equation}
This can be written as
\begin{equation}\label{eq:FDdefinition2}
  \F(x,\pbar)=1+\frac{\ksl}{2\pbar{\cdot}k}(\Vsl\ee^{-i\kx}+\Vssl\ee^{i\kx})\,,
\end{equation}
where we define $\Vssl=\laser^{*\mu}\gamma_\mu$, that is  we do not take the complex conjugate of the Dirac matrices.

In addition, from (\ref{eq:fvolkov}) and (\ref{eq:fbarvolkov}), we see that the laser background distorts the plane waves of the free theory which, in this interacting theory, become
\begin{align}\label{eq:Ddefinition}
\begin{split}
  \De(x,\pbar)&=\ee^{-i\pbar{\cdot}x}\ee^{i(\bar{\omega}_1\sin(\kdotx)+\bar{v}\sin(2\kdotx)+\bar{\omega}_2\cos(\kdotx))} 
  \\
  &=\ee^{-i\pbar{\cdot}x}\sum_{n=-\infty}^\infty \ee^{in \kdotx}\J_n(\pbar)\,,
 \end{split}  
\end{align}
where we have introduced $\J_n(\pbar)$, an elliptical generalisation of the Bessel functions, which we call eccentric Bessel functions. Their properties are discussed in Appendix~\ref{sec:appen2}. The variables $\bar{\omega}_1$, $\bar{\omega}_2$  and $\bar{v}$ in the upper line of~(\ref{eq:Ddefinition}) are given by
\begin{equation}\label{eq:w1}
  \bar{\omega}_1=-\left( \frac{\pbar{\cdot}\laser ^*}{\pbar{\cdot}k}+\frac{\pbar{\cdot}\laser}{\pbar{\cdot}k}\right)\,,\qquad
  \bar{\omega}_2=i\left( \frac{\pbar{\cdot}\laser^*}{\pbar{\cdot}k}-\frac{\pbar{\cdot}\laser}{\pbar{\cdot}k}\right)
\end{equation}
and
\begin{equation}\label{eq:v}
  \bar{v}=\left(\frac{{\laser^*}^2}{4\pbar{\cdot}k}+\frac{{\laser}^2}{4\pbar{\cdot}k}\right)\,.
\end{equation}
The bar over these variables indicates their construction out of the on-shell momentum $\pbar$, as in~(\ref{eq:on_shell}). Later, when this momentum $\pbar$ is extended off shell and written as $p$, all these variables will also lose their bars.

Note that in the limit of linear polarisation, where $\laser^\mu$ is real,  we have 
\begin{equation}
  \bar{\omega}_1\big|_{\ecc=1}=-2\frac{\pbar{\cdot}\laser}{\pbar{\cdot}k}\,,\qquad
  \bar{\omega}_2\big|_{\ecc=1}=0\,,\qquad 
  \bar{v}\big|_{\ecc=1}=\frac{{\laser}^2}{2\pbar{\cdot}k}\,,
\end{equation}
while for circular polarisation we have 
\begin{equation}
  \bar{\omega}_1\big|_{\ecc=0}=-\left( \frac{\pbar{\cdot}\laser^*}{\pbar{\cdot}k}+\frac{\pbar{\cdot}\laser}{\pbar{\cdot}k}\right)\,,\qquad
  \bar{\omega}_2\big|_{\ecc=0}=i\left( \frac{\pbar{\cdot}\laser^*}{\pbar{\cdot}k}-\frac{\pbar{\cdot}\laser}{\pbar{\cdot}k}\right)\,,\qquad
  \bar{v}\big|_{\ecc=0}=0\,.
\end{equation}
In terms of the original amplitudes, this circular limit has the simple form
\begin{equation}
  \bar{\omega}_1\big|_{\ecc=0}=-e\frac{p{\cdot}a_1}{p{\cdot}k}\,,\qquad
  \bar{\omega}_2\big|_{\ecc=0}=e\frac{p{\cdot}a_2}{p{\cdot}k}\,.
\end{equation}

It is important to note that there is a residual gauge freedom present in our description of the background field (\ref{eq:Aa})  given by $\Ac_\mu(x)\to \Ac_\mu(x)+\partial_\mu \lambda(x)$ where
\begin{equation}
  \lambda(x)=\lambda_1\eccp\sin(k{\cdot}x)-\lambda_2 \eccm\cos(k{\cdot}x)\,.
\end{equation}
Transformations of this form preserve the elliptic class and act on the amplitudes to give
\begin{equation}
  a_1^\mu\to a_1^\mu+\lambda_1 k^\mu\,,\qquad a_2^\mu\to a_2^\mu+\lambda_2 k^\mu\,.
\end{equation}
Through this the undetermined parameters in (\ref{eq:a1a2}) can be eliminated, thus simplifying the route to the Stokes' parameters. 

In terms of the complex amplitude $\laser^\mu$ these gauge transformations become
\begin{equation}
  \laser^\mu\to \laser^\mu+\tfrac12 e(\lambda_1\eccp+i \lambda_2\eccm)k^\mu\,,
\end{equation}
 and we see  that both $\laser^2$ and $\laser{\cdot}\laser^*$ are invariant under this residual gauge freedom.  Hence the mass shift 
   (\ref{eq:mass_star_square}) is also gauge invariant. 
Furthermore, from (\ref{eq:w1}) and (\ref{eq:v}),  we have $\omega_1\to \omega_1-e \lambda_1\eccp$,  $\omega_2\to \omega_2+e \lambda_2\eccm$ and $v\to v$. Tracing these changes through (\ref{eq:Ddefinition}) and (\ref{eq:fvolkov})  then leads to the expected background field  residual gauge transformation property within this elliptic class: 
\begin{equation}
  \psiV(x)\to \ee^{-ie\lambda(x)}\psiV(x)\,.
\end{equation}

In this section we have presented the solution to the Dirac equation in an elliptically polarised plane wave background, and we have seen the polarisation independence and residual gauge invariance of the various manifestations of the fermionic  mass shift. Next we will turn to constructing the propagator and, more generally, the two point function for an electron in such a  background.

\section{The Two Point Function}\label{sec:twopoint} 
The Volkov field for the elliptic class of polarisations can now be used to construct the two point function through the (Volkov) vacuum expectation values of the  time-ordered product of such fields.  Extending  the arguments given in \cite{Lavelle:2015jxa} for the linearly polarised laser, we find for the full two point function
\begin{align}\label{eq:two_point_volkov_con}
\begin{split}
  \braV{0}T\psiV(x)\psibV(y)\ketV{0}&=\sum_{n,r=-\infty}^\infty\ee^{in\kx}\ee^{-ir\ky}\intfp\ee^{-ip{\cdot}(x-y)}\\&\qquad\times
  \sqrt{\vphantom{1^1}{\boldsymbol{Z}}_2(n,p)} \frac{i(\psl+m- \delta\msl)}{p^2-m_*^2+i\epsilon}\sqrt{\vphantom{1^1}\smash{\overline{\boldsymbol{Z}}}_2(r,p)}\,,
  \end{split}
\end{align} 
In this expression we see the expected normalised fermionic propagator with off-shell momentum $p^\mu$ but with several additional structures. 

The free propagator-like term in (\ref{eq:two_point_volkov_con}) is enclosed between wave function normalisation factors, constructed out of the eccentric Bessel functions introduced in Appendix~\ref{sec:appen2},  given by
 \begin{equation}\label{eq:Z2}
   \sqrt{\vphantom{1^1}{\boldsymbol{Z}}_2(n,p)}=\J_n(p)+\frac{\ksl}{2p{\cdot}k}\big(\Vsl \J_{n+1}(p)+\Vsl^{*}\J_{n-1}(p)\big)
 \end{equation}
and
\begin{equation}\label{eq:Z2bar}
 \sqrt{\vphantom{1^1}\smash{\overline{\boldsymbol{Z}}}_2(r,p)}= \Js_r(p)-\frac{\ksl}{2p{\cdot}k}\big(\Vsl^{*}\Js_{r+1}(p)+\Vsl \Js_{r-1}(p)\big)\,,
\end{equation}
which is the adjoint normalisation factor defined via
\begin{equation}
   \sqrt{\vphantom{1^1}\smash{\overline{\boldsymbol{Z}}}_2(r,p)}=\gamma_0 \sqrt{\vphantom{1^1}{\boldsymbol{Z}}_2(r,p)} ^\dag \gamma_0\,.
 \end{equation}

The next  thing to notice is that translational invariance  has been partially lost in~(\ref{eq:two_point_volkov_con}) due to the phase factors in front of the integral. This  obstructs a fully momentum space description of the two point function. This is not a surprise and  simply reflects the existence of the fixed laser background. What is interesting to note, though, is that a fully momentum space description can be recovered if $k{\cdot}x$ or $k{\cdot}y$ vanishes. Also, the diagonal part of the two point function, where $n=r$, is fully translational invariant. 

Finally, the two point function  (\ref{eq:two_point_volkov_con}) contains a matrix mass shift given by 
\begin{equation}\label{eq:mms}
  \delta\msl=-\frac{\laser{\cdot}\laser^*}{p{\cdot}k}\ksl=-\frac{e^2a^2}{2p{\cdot}k}\ksl\,.
\end{equation}
As well as the unusual feature of this being a matrix~\cite{Lavelle:2015jxa}, the relative sign in the numerator,  $m-\delta\msl$, looks surprising. It can be understood in this fermionic theory by rewriting
\begin{align}
\label{eq:signcaution}
  \frac{i(\psl+m- \delta\msl)}{p^2-m^2_*}=\frac{i}{\psl-({}m+\delta\msl)}\,,
\end{align}
where we have used (\ref{eq:mass_star_square}). We emphasise again that this shows that both manifestations (\ref{eq:mass_star_square}) and (\ref{eq:mms}) of the mass shift in this fermionic theory  are gauge invariant and independent of the polarisation within this elliptic class of backgrounds.

Our expression (\ref{eq:two_point_volkov_con}) for the two point function  explicitly brings out the mass shift caused by the laser background. 
Other approaches to the Volkov field, for specific polarisations,  write the two point function in different ways and we will now see how to make contact with such approaches for the full elliptic class. 

It is possible to  rewrite (\ref{eq:two_point_volkov_con}) in a more familiar form where the mass shift is, though, not manifest.
To do this, shift the momentum in (\ref{eq:two_point_volkov_con}) by 
\begin{equation*}
  p^\mu\to p^\mu-\frac{\laser^*{\cdot}\laser}{p{\cdot}k}k^\mu\,.
\end{equation*}
This leads to the following expression  for the two point function
\begin{align}\label{eq:two_point_volkov_E}
  \braV{0}T\psiV(x)\psibV(y)\ketV{0}&=\intfp E(x,p) \frac{i(\psl+m)}{p^2-m^2+i\epsilon}\bar{E}(y,p)\,,
\end{align}
where 
\begin{equation}
  E(x,p)=\ee^{-ip{\cdot}x}\sum_{n=-\infty}^\infty \ee^{i(n+\frac{\laser^*{\cdot}\laser}{p{\cdot}k})k{\cdot}x}\sqrt{\vphantom{1^1}{\boldsymbol{Z}}_2(n,p)}\,,
\end{equation}
and the mass shift has been absorbed into the phase.

The factor, $E(x,p)$, may also be rewritten by  inserting expressions (\ref{eq:Z2}) and (\ref{eq:Z2bar}) for the normalisation factors, and then, through shifts in the sum, extracting an overall eccentric Bessel function to get
\begin{equation}
  E(p,x)=\ee^{-ip{\cdot}x}\sum_{n=-\infty}^\infty \ee^{ink{\cdot}x}\J_n(p)
  \Big(
  1+\frac{\ksl}{2p{\cdot}k}\left(
  \Vsl\ee^{-ik{\cdot}x} + \Vsl^*\ee^{ik{\cdot}x}
  \right)
  \Big)\ee^{i\frac{\laser^*{\cdot}\laser}{p{\cdot}k}k{\cdot}x}\,.
\end{equation}
From (\ref{eq:Fdefinition}), (\ref{eq:FDdefinition2}) and our definition (\ref{eq:eccentricdef}) of the eccentric Bessel functions, this becomes
\begin{equation}
  E(p,x)=\ee^{-i(p-\frac{\laser^*{\cdot}\laser}{p{\cdot}k}k){\cdot}x}
  \Big(
  1+\frac{e\ksl\Asl(k{\cdot}x)}{2p{\cdot}k}\Big)
  \ee^{i(\omega_1\sin(k{\cdot} x)+\omega_2\cos(k{\cdot} x)+v\sin(2k{\cdot} x))}\,.
\end{equation}
This then rapidly leads to the expression that
\begin{equation}\label{eq:ritus}
  E(p,x)=
  \Big(
  1+\frac{e\ksl\Asl(k{\cdot}x)}{2p{\cdot}k}\Big)
  \ee^{iS(p,x)}\,,
\end{equation}
where the phase factor includes integrals over the background potential
\begin{equation}\label{eq:sfactor}
  S(p,x)=-p{\cdot}x-\int^{k{\cdot}x}\!\!\! d\phi 
  \left(\frac{ep{\cdot}\Ac(\phi)}{p{\cdot}k}-
  \frac{e^2\Ac^2(\phi)}{2p{\cdot}k}\right)\,.
\end{equation}
The formal expression (\ref{eq:ritus})  for the Ritus matrices~\cite{Ritus:1972ky} assumes that there are appropriate boundary conditions  defined  so that  the l{}ower limit in the integral~(\ref{eq:sfactor}) does  not contribute.

An alternative way to rewrite  (\ref{eq:two_point_volkov_con}) is to focus on the factors leading to the  break down in translational invariance, and try to minimise their impact on the overall structure of the two point function. The pre-factor in (\ref{eq:two_point_volkov_con}) responsible for this violation  can be written in several equivalent ways. For example, we could have
 \begin{equation}
   \ee^{in\kx}\ee^{-ir\ky}=\ee^{ink{\cdot}(x-y)}\ee^{i(n-r)\ky}
 \end{equation}
 or 
 \begin{equation}
   \ee^{in\kx}\ee^{-ir\ky}=\ee^{i(n-r)k{\cdot}x}\ee^{irk{\cdot}(x-y)}\,.
 \end{equation}
 The translation invariant part of either of these can then be absorbed into the integral by shifting either $p\to p+nk$ or $p\to p+rk$ and redefining the dummy variables in the sum. This leads to two alternative but complementary ways for writing (\ref{eq:two_point_volkov_con}) as either
\begin{align}\label{eq:two_pint_volkov_v3}
\begin{split}
  \braV{0}T\psiV(x)\psibV(y)\ketV{0}&=\sum_{r=-\infty}^\infty\ee^{irk{\cdot}y}\intfp\ee^{-ip{\cdot}(x-y)}\\&\qquad\times \sum_{n=-\infty}^\infty
  \sqrt{\vphantom{1^1}{\boldsymbol{Z}}_2(n,p)} \frac{i(\psl+n\ksl+m- \delta\msl)}{(p+nk)^2-m_*^2+i\epsilon}\sqrt{\vphantom{1^1}\smash{\overline{{\boldsymbol{Z}}}}_2(n-r,p)}
  \end{split}
\end{align}
or
\begin{align}\label{eq:two_pint_volkov_v4}
\begin{split}
  \braV{0}T\psiV(x)\psibV(y)\ketV{0}&=\sum_{r=-\infty}^\infty\ee^{ir\kx}\intfp\ee^{-ip{\cdot}(x-y)}\\&\qquad\times \sum_{n=-\infty}^\infty
  \sqrt{\vphantom{1^1}{\boldsymbol{Z}}_2(n+r,p)} \frac{i(\psl+n\ksl+m- \delta\msl)}{(p+nk)^2-m_*^2+i\epsilon}\sqrt{\vphantom{1^1}\smash{\overline{{\boldsymbol{Z}}}}_2(n,p)}\,.   
  \end{split}
\end{align}
In contrast to (\ref{eq:two_point_volkov_con}), in both of these expressions for the two point function, the central propagator like structure now depends on one of  the momentum  summations. This exposes the sum over sideband poles interpretation of  the two point function seen earlier for specific polarisations~\cite{Eberly:1966b}. This description in terms of sidebands naturally emerges in perturbative calculations in weak laser background~\cite{Lavelle:2013wx}, \cite{Lavelle:2015jxa}, and  thus offers an attractive route to the introduction of loop corrections and the associated renormalisation process.

\section{Conclusions} 
In this paper we have seen how to formulate a description of a fermion propagating in an elliptical class of laser backgrounds. This class  includes both linear and circular polarisations as particular limits of the eccentricity parameter. Through this we have seen how to identify eccentricity dependences  in the structure of the Volkov field and its two point function. 

An unexpected result that emerges from working in this full class of polarisations is that the (matrix) mass shift is independent of the eccentricity of the background field. Different values of the mass shift in the literature are understood as arising from different choices of normalisation of the background field. We have seen how this normalisation can be motivated physically in terms of the overall energy of the system, which is conserved within this elliptic class.

Although the presence of any laser field violates translational invariance, we have seen that it is still possible to develop a momentum space description of the two point function in conjunction with phase factors that contain the translation non-invariant physics. 

\section{Acknowledgements} 
We thank Tom Heinzl, Anton Ilderton and Ben King for discussions and comments.  We also thank the referee for helpful comments.
\appendix
{} 

\section{Volkov field details}\label{sec:appen1}
In this appendix we will sketch how the solution of the  Dirac equation  (\ref{eq:eqm}) in an elliptically polarised plane wave background is obtained. 

From (\ref{eq:fvolkov}) we see that the derivative acts upon the terms $\F(x,\pbar)\De(x,\pbar)$. Using (\ref{eq:FDdefinition2}) and (\ref{eq:Ddefinition}) we obtain
\begin{equation}
  i(\dsl+ie\Asl(x))\F(x,\pbar)\De(x,\pbar)=\F(x,\pbar)\De(x,\pbar)(\pbsl+\frac{\laser{\cdot}\laser^*}{\pbar{\cdot}k}\ksl)
\end{equation}
and equivalently
\begin{equation}
  i(\dsl+ie\Asl(x))\F(x,-\pbar)\De(x,-\pbar)=\F(x,-\pbar)\De(x,-\pbar)(-\pbsl-\frac{\laser{\cdot}\laser^*}{\pbar{\cdot}k}\ksl)\,.
\end{equation}
Note that in both of these equations we see that the action of the Dirac equation has been pulled through to now act upon the spinors in~(\ref{eq:fvolkov}).

Then, in order for (\ref{eq:eqm}) to be solved, acting on the spinors we need 
\begin{equation}\label{eq:uvolkov}
  (\pbsl+\frac{\laser{\cdot}\laser^*}{\pbar{\cdot}k}\ksl)\Uv^{(\alpha)}(p)=m\,\Uv^{(\alpha)}(p)\,,
\end{equation}
and
\begin{equation}\label{eq:vvolkov}
  (\pbsl+\frac{\laser{\cdot}\laser^*}{\pbar{\cdot}k}\ksl)\Vv^{(\alpha)}(p)=-m\,\Vv^{(\alpha)}(p)\,.
\end{equation}
From (\ref{eq:vvstar}) these equations are independent of the eccentricity of the polarisation.

We can now  construct the Volkov spinors by boosting the static spinors
\begin{equation}
    \U^{(1)}(0)=\begin{pmatrix}
      1\\0\\0\\0
    \end{pmatrix},\quad
    \U^{(2)}(0)=\begin{pmatrix}
      0\\1\\0\\0
    \end{pmatrix},\quad
    \V^{(1)}(0)=\begin{pmatrix}
      0\\0\\1\\0
    \end{pmatrix},\quad
    \V^{(2)}(0)=\begin{pmatrix}
      0\\0\\0\\1
    \end{pmatrix}.
  \end{equation}
We define 
\begin{equation}
  \Uv^{(\alpha)}(p)=\frac1{\sqrt{2m(m+E^*_p+\frac{\laser{\cdot}\laser^*}{\pbar{\cdot}k} k_0)}}\bigg(\pbsl+\frac{\laser{\cdot}\laser^*}{\pbar{\cdot}k}\ksl+m\bigg)\U^{(\alpha)}(0)\,,
\end{equation}
and
\begin{equation}
  \Vv^{(\alpha)}(p)=\frac1{\sqrt{2m(m+E^*_p+\frac{\laser{\cdot}\laser^*}{\pbar{\cdot}k} k_0)}}\bigg(-\pbsl-\frac{\laser{\cdot}\laser^*}{\pbar{\cdot}k}\ksl+m\bigg)\V^{(\alpha)}(0)\,.
\end{equation}
With these conventions we  have the inner products 
\begin{equation}
  \Uvb^{(\alpha)}(p)\Uv^{(\beta)}(p)=\delta^{\alpha \beta}\,,\quad\mbox{and}\quad
  \Vvb^{(\alpha)}(p)\Vv^{(\beta)}(p)=-\delta^{\alpha \beta}\,,
\end{equation}
as well as the tensor products
\begin{equation}
  \Uv^{(\alpha)}(p)\Uvb^{(\alpha)}(p)=\frac{\pbsl +\frac{\laser{\cdot}\laser^*}{\pbar{\cdot}k}\ksl+m}{2m}\,,
\end{equation}
and 
\begin{equation}
  \Vv^{(\alpha)}(p)\Vvb^{(\alpha)}(p)=\frac{\pbsl +\frac{\laser{\cdot}\laser^*}{\pbar{\cdot}k}\ksl-m}{2m}\,.
\end{equation}

The Volkov Fock states are then built up in the usual way from the  Volkov vacuum $\ketV{0}$. Note that with our conventions,  the creation-annihilation operators satisfy the anti-commutation relation 
\begin{equation}
  \{ \aVa(p),\adVb(q)\}
  =
  (2\pi)^3\frac{E_p^*}m  \delta^{\alpha\beta} 
  \delta^{(3)}(p-q)\,.
\end{equation}

The two point function is the time-ordered product of the vacuum expectation value of the fields $\psiV(x)$ and $\psibV(y)$. This means that we first need to evaluate $\braV{0}\psiV(x)\psibV(y)\ketV{0}$ and $\braV{0}\psibV(y)\psiV(x)\ketV{0}$, and secondly to impose the time-ordering. A little algebra shows that
\begin{equation}\label{eq:vac1} 
  \braV{0}\psiV(x)\psibV(y)\ket{0}=\intfps\frac{1}{2E^{*}_p}\F(x,\pbar)\De(x,\pbar)(\pbsl+m- \delta\mslb)\F(y,-\pbar)\De^\dag(y,\pbar)
\end{equation} 
and
\begin{equation}\label{eq:vac2}
  \braV{0}\psibV(y)\psiV(x)\ketV{0}=\intfps\frac{1}{2E^{*}_p}\F(x,-\pbar)\De(x,-\pbar)(\pbsl-m- \delta\mslb)\F(y,\pbar)\De^\dag(y,-\pbar)\,.
\end{equation}
We now need to tidy these expressions up prior to inserting the time-ordering. Working through the definitions we  see that
\begin{equation}
  \F(x,\pbar)\De(x,\pbar)=\ee^{-i\pbar{\cdot}x}\sum_{n}\ee^{in\kx}\Big(\J_n(\pbar)+\frac{\ksl}{2\pbar{\cdot}k}(\Vsl \J_{n+1}(\pbar)+\Vsl^{*}\J_{n-1}(\pbar))\Big)
\end{equation}
and 
\begin{equation}
  \F(y,-\pbar)\De^\dag(y,\pbar)=\ee^{i\pbar{\cdot}y}\sum_{r}\ee^{-ir\kx}\Big(\Js_r(\pbar)-\frac{\ksl}{2\pbar{\cdot}k}(\Vsl \Js_{r-1}(\pbar)+\Vsl^{*}\Js_{r+1}(\pbar))\Big)\,,
\end{equation}
where we have introduced the elliptic Bessel functions of Appendix~B.

Hence we find that 
\begin{align}
\begin{split}
  \braV{0}\psiV(x)\psibV(y)\ketV{0}&=\sum_{n,r}\ee^{in\kx}\ee^{-ir\ky}\intfps\frac{1}{2E^{*}_p}\ee^{-i\pbar{\cdot}(x-y)}\\&\qquad\times
  \Big(\J_n(\pbar)+\frac{\ksl}{2\pbar{\cdot}k}\big(\Vsl \J_{n+1}(\pbar)+\Vsl^{*}\J_{n-1}(\pbar)\big)\Big)\\&\qquad\quad\times(\pbsl+m- \delta\mslb)
  \\&\qquad\qquad\times\Big(\Js_r(\pbar)-\frac{\ksl}{2\pbar{\cdot}k}\big(\Vsl \Js_{r-1}(\pbar)+\Vsl^{*}\Js_{r+1}(\pbar)\big)\Big)
  \end{split}
\end{align}
and    
\begin{align}
\begin{split}
  \braV{0}\psibV(y)\psiV(x)\ketV{0}&=\sum_{n,r}\ee^{in\kx}\ee^{-ir\ky}\intfps\frac{1}{2E^{*}_p}\ee^{i\pbar{\cdot}(x-y)}\\&\qquad\times
  \Big(\J_n(-\pbar)-\frac{\ksl}{2\pbar{\cdot}k}\big(\Vsl \J_{n+1}(-\pbar)+\Vsl^{*}\J_{n-1}(-\pbar)\big)\Big)\\&\qquad\quad\times(\pbsl-m- \delta\mslb)
  \\&\qquad\qquad\times\Big(\Js_r(-\pbar)+\frac{\ksl}{2\pbar{\cdot}k}\big(\Vsl \Js_{r-1}(-\pbar)+\Vsl^{*}\Js_{r+1}(-\pbar)\big)\Big)\,.
  \end{split}
\end{align}
If we now add the time ordering and follow the same steps as in~\cite{Lavelle:2015jxa}, the momentum is extended off-shell (so $\pbar\to p$) and we obtain~(\ref{eq:two_point_volkov_con}).

\section{Eccentric Bessel Functions} \label{sec:appen2}

Recall that the  Bessel functions (of the first kind) are defined by
\begin{equation}
  \ee^{ia\sin\theta}=\sum_{n=-\infty}^\infty \ee^{in\theta}J_n(a)
 \end{equation} 
A similar expansion with $\cos\theta$ in the exponential is easy to derive and we get
\begin{equation}
  \ee^{ic\cos\theta}=\ee^{ic\sin(\theta+\frac{\pi}2)}=\sum_{m=-\infty}^\infty \ee^{im\theta}i^mJ_m(c)
 \end{equation} 
Using these we see that
\begin{align*} 
  \ee^{ia\sin\theta}\ee^{ib\sin2\theta}\ee^{ic\cos\theta}
  =\sum_{n}\ee^{in\theta}\sum_{m} i^mJ_{n-m}(a,b)J_m(c)
\end{align*} 
where we have introduced the generalised Bessel functions~\cite{Krainov:1997}\cite{Lavelle:2013wx}, familiar from  studies of  linearly polarised lasers, defined by
\begin{equation}
  J_n(a,b)=\sum_r J_{n-2r}(a)J_r(b)\,.
\end{equation}
We now define the \emph{eccentric Bessel functions} by
\begin{equation}
  J_n(a,b,c)=\sum_mi^m J_{n-m}(a,b)J_m(c)\,,
\end{equation}
then we have
\begin{equation}\label{eq:eccentricdef}
  \ee^{ia\sin\theta}\ee^{ib\sin2\theta}\ee^{ic\cos\theta}=\sum_n\ee^{in\theta}J_n(a,b,c)=\sum_n\ee^{in\theta}\J_n(p)\,,
\end{equation}
where the final expression uses the  condensed notation used in the main part of this paper.



\begin{thebibliography}{30}%
\makeatletter
\providecommand \@ifxundefined [1]{%
 \@ifx{#1\undefined}
}%
\providecommand \@ifnum [1]{%
 \ifnum #1\expandafter \@firstoftwo
 \else \expandafter \@secondoftwo
 \fi
}%
\providecommand \@ifx [1]{%
 \ifx #1\expandafter \@firstoftwo
 \else \expandafter \@secondoftwo
 \fi
}%
\providecommand \natexlab [1]{#1}%
\providecommand \enquote  [1]{``#1''}%
\providecommand \bibnamefont  [1]{#1}%
\providecommand \bibfnamefont [1]{#1}%
\providecommand \citenamefont [1]{#1}%
\providecommand \href@noop [0]{\@secondoftwo}%
\providecommand \href [0]{\begingroup \@sanitize@url \@href}%
\providecommand \@href[1]{\@@startlink{#1}\@@href}%
\providecommand \@@href[1]{\endgroup#1\@@endlink}%
\providecommand \@sanitize@url [0]{\catcode `\\12\catcode `\$12\catcode
  `\&12\catcode `\#12\catcode `\^12\catcode `\_12\catcode `\%12\relax}%
\providecommand \@@startlink[1]{}%
\providecommand \@@endlink[0]{}%
\providecommand \url  [0]{\begingroup\@sanitize@url \@url }%
\providecommand \@url [1]{\endgroup\@href {#1}{\urlprefix }}%
\providecommand \urlprefix  [0]{URL }%
\providecommand \Eprint [0]{\href }%
\providecommand \doibase [0]{http://dx.doi.org/}%
\providecommand \selectlanguage [0]{\@gobble}%
\providecommand \bibinfo  [0]{\@secondoftwo}%
\providecommand \bibfield  [0]{\@secondoftwo}%
\providecommand \translation [1]{[#1]}%
\providecommand \BibitemOpen [0]{}%
\providecommand \bibitemStop [0]{}%
\providecommand \bibitemNoStop [0]{.\EOS\space}%
\providecommand \EOS [0]{\spacefactor3000\relax}%
\providecommand \BibitemShut  [1]{\csname bibitem#1\endcsname}%
\let\auto@bib@innerbib\@empty
\bibitem [{\citenamefont {Heinzl}(2012)}]{Heinzl:2011ur}%
  \BibitemOpen
  \bibfield  {author} {\bibinfo {author} {\bibfnamefont {T.}~\bibnamefont
  {Heinzl}},\ }\href {\doibase 10.1142/S2010194512007283,
  10.1142/S0217751X1260010X} {\bibfield  {journal} {\bibinfo  {journal} {Int.
  J. Mod. Phys.}\ }\textbf {\bibinfo {volume} {A27}},\ \bibinfo {pages}
  {1260010} (\bibinfo {year} {2012})},\ \Eprint
  {http://arxiv.org/abs/1111.5192} {arXiv:1111.5192 [hep-ph]} \BibitemShut
  {NoStop}%
\bibitem [{\citenamefont {Di~Piazza}\ \emph {et~al.}(2012)\citenamefont
  {Di~Piazza}, \citenamefont {Muller}, \citenamefont {Hatsagortsyan},\ and\
  \citenamefont {Keitel}}]{DiPiazza:2011tq}%
  \BibitemOpen
  \bibfield  {author} {\bibinfo {author} {\bibfnamefont {A.}~\bibnamefont
  {Di~Piazza}}, \bibinfo {author} {\bibfnamefont {C.}~\bibnamefont {Muller}},
  \bibinfo {author} {\bibfnamefont {K.}~\bibnamefont {Hatsagortsyan}}, \ and\
  \bibinfo {author} {\bibfnamefont {C.}~\bibnamefont {Keitel}},\ }\href
  {\doibase 10.1103/RevModPhys.84.1177} {\bibfield  {journal} {\bibinfo
  {journal} {Rev. Mod. Phys.}\ }\textbf {\bibinfo {volume} {84}},\ \bibinfo
  {pages} {1177} (\bibinfo {year} {2012})},\ \Eprint
  {http://arxiv.org/abs/1111.3886} {arXiv:1111.3886 [hep-ph]} \BibitemShut
  {NoStop}%
\bibitem [{\citenamefont {Volkov}(1935)}]{Volkov:1935zz}%
  \BibitemOpen
  \bibfield  {author} {\bibinfo {author} {\bibfnamefont {D.~M.}\ \bibnamefont
  {Volkov}},\ }\href {\doibase 10.1007/BF01331022} {\bibfield  {journal}
  {\bibinfo  {journal} {Z. Phys.}\ }\textbf {\bibinfo {volume} {94}},\ \bibinfo
  {pages} {250} (\bibinfo {year} {1935})}\BibitemShut {NoStop}%
\bibitem [{\citenamefont {Reiss}\ and\ \citenamefont
  {Eberly}(1966)}]{Reiss:1966A}%
  \BibitemOpen
  \bibfield  {author} {\bibinfo {author} {\bibfnamefont {H.~R.}\ \bibnamefont
  {Reiss}}\ and\ \bibinfo {author} {\bibfnamefont {J.~H.}\ \bibnamefont
  {Eberly}},\ }\href {\doibase 10.1103/PhysRev.151.1058} {\bibfield  {journal}
  {\bibinfo  {journal} {Phys. Rev.}\ }\textbf {\bibinfo {volume} {151}},\
  \bibinfo {pages} {1058} (\bibinfo {year} {1966})}\BibitemShut {NoStop}%
\bibitem [{\citenamefont {Brown}\ and\ \citenamefont
  {Kibble}(1964)}]{Brown:1964zz}%
  \BibitemOpen
  \bibfield  {author} {\bibinfo {author} {\bibfnamefont {L.~S.}\ \bibnamefont
  {Brown}}\ and\ \bibinfo {author} {\bibfnamefont {T.~W.~B.}\ \bibnamefont
  {Kibble}},\ }\href {\doibase 10.1103/PhysRev.133.A705} {\bibfield  {journal}
  {\bibinfo  {journal} {Phys. Rev. A}\ }\textbf {\bibinfo {volume} {133}},\
  \bibinfo {pages} {705} (\bibinfo {year} {1964})}\BibitemShut {NoStop}%
\bibitem [{\citenamefont {Neville}\ and\ \citenamefont
  {Rohrlich}(1971)}]{Neville:1971uc}%
  \BibitemOpen
  \bibfield  {author} {\bibinfo {author} {\bibfnamefont {R.~A.}\ \bibnamefont
  {Neville}}\ and\ \bibinfo {author} {\bibfnamefont {F.}~\bibnamefont
  {Rohrlich}},\ }\href {\doibase 10.1103/PhysRevD.3.1692} {\bibfield  {journal}
  {\bibinfo  {journal} {Phys. Rev.}\ }\textbf {\bibinfo {volume} {D3}},\
  \bibinfo {pages} {1692} (\bibinfo {year} {1971})}\BibitemShut {NoStop}%
\bibitem [{\citenamefont {Dittrich}(1972{\natexlab{a}})}]{Dittrich:1973rn}%
  \BibitemOpen
  \bibfield  {author} {\bibinfo {author} {\bibfnamefont {W.}~\bibnamefont
  {Dittrich}},\ }\href {\doibase 10.1103/PhysRevD.6.2094} {\bibfield  {journal}
  {\bibinfo  {journal} {Phys. Rev.}\ }\textbf {\bibinfo {volume} {D6}},\
  \bibinfo {pages} {2094} (\bibinfo {year} {1972}{\natexlab{a}})}\BibitemShut
  {NoStop}%
\bibitem [{\citenamefont {Dittrich}(1972{\natexlab{b}})}]{Dittrich:1973rm}%
  \BibitemOpen
  \bibfield  {author} {\bibinfo {author} {\bibfnamefont {W.}~\bibnamefont
  {Dittrich}},\ }\href {\doibase 10.1103/PhysRevD.6.2104} {\bibfield  {journal}
  {\bibinfo  {journal} {Phys. Rev.}\ }\textbf {\bibinfo {volume} {D6}},\
  \bibinfo {pages} {2104} (\bibinfo {year} {1972}{\natexlab{b}})}\BibitemShut
  {NoStop}%
\bibitem [{\citenamefont {Kibble}\ \emph {et~al.}(1975)\citenamefont {Kibble},
  \citenamefont {Salam},\ and\ \citenamefont {Strathdee}}]{Kibble:1975vz}%
  \BibitemOpen
  \bibfield  {author} {\bibinfo {author} {\bibfnamefont {T.~W.~B.}\
  \bibnamefont {Kibble}}, \bibinfo {author} {\bibfnamefont {A.}~\bibnamefont
  {Salam}}, \ and\ \bibinfo {author} {\bibfnamefont {J.~A.}\ \bibnamefont
  {Strathdee}},\ }\href {\doibase 10.1016/0550-3213(75)90581-7} {\bibfield
  {journal} {\bibinfo  {journal} {Nucl. Phys.}\ }\textbf {\bibinfo {volume}
  {B96}},\ \bibinfo {pages} {255} (\bibinfo {year} {1975})}\BibitemShut
  {NoStop}%
\bibitem [{\citenamefont {Mitter}(1975)}]{Mitter:1974yg}%
  \BibitemOpen
  \bibfield  {author} {\bibinfo {author} {\bibfnamefont {H.}~\bibnamefont
  {Mitter}},\ }\href {\doibase 10.1007/978-3-7091-8424-0_7} {\bibfield
  {journal} {\bibinfo  {journal} {Acta Phys. Austriaca Suppl.}\ }\textbf
  {\bibinfo {volume} {14}},\ \bibinfo {pages} {397} (\bibinfo {year}
  {1975})}\BibitemShut {NoStop}%
\bibitem [{\citenamefont {Ritus}(1985)}]{Ritus:1985review}%
  \BibitemOpen
  \bibfield  {author} {\bibinfo {author} {\bibfnamefont {V.~I.}\ \bibnamefont
  {Ritus}},\ }\href {\doibase 10.1007/BF01120220} {\bibfield  {journal}
  {\bibinfo  {journal} {J Russ Laser Res}\ }\textbf {\bibinfo {volume} {6}},\
  \bibinfo {pages} {497} (\bibinfo {year} {1985})}\BibitemShut {NoStop}%
\bibitem [{\citenamefont {Ilderton}\ and\ \citenamefont
  {Torgrimsson}(2013)}]{Ilderton:2012qe}%
  \BibitemOpen
  \bibfield  {author} {\bibinfo {author} {\bibfnamefont {A.}~\bibnamefont
  {Ilderton}}\ and\ \bibinfo {author} {\bibfnamefont {G.}~\bibnamefont
  {Torgrimsson}},\ }\href {\doibase 10.1103/PhysRevD.87.085040} {\bibfield
  {journal} {\bibinfo  {journal} {Phys. Rev.}\ }\textbf {\bibinfo {volume}
  {D87}},\ \bibinfo {pages} {085040} (\bibinfo {year} {2013})},\ \Eprint
  {http://arxiv.org/abs/1210.6840} {arXiv:1210.6840 [hep-th]} \BibitemShut
  {NoStop}%
\bibitem [{\citenamefont {Lavelle}\ \emph {et~al.}(2013)\citenamefont
  {Lavelle}, \citenamefont {McMullan},\ and\ \citenamefont
  {Raddadi}}]{Lavelle:2013wx}%
  \BibitemOpen
  \bibfield  {author} {\bibinfo {author} {\bibfnamefont {M.}~\bibnamefont
  {Lavelle}}, \bibinfo {author} {\bibfnamefont {D.}~\bibnamefont {McMullan}}, \
  and\ \bibinfo {author} {\bibfnamefont {M.}~\bibnamefont {Raddadi}},\ }\href
  {\doibase 10.1103/PhysRevD.87.085024} {\bibfield  {journal} {\bibinfo
  {journal} {Phys. Rev.}\ }\textbf {\bibinfo {volume} {D87}},\ \bibinfo {pages}
  {085024} (\bibinfo {year} {2013})},\ \Eprint {http://arxiv.org/abs/1301.3072}
  {arXiv:1301.3072 [hep-ph]} \BibitemShut {NoStop}%
\bibitem [{\citenamefont {Lavelle}\ and\ \citenamefont
  {McMullan}(2014)}]{Lavelle:2014mka}%
  \BibitemOpen
  \bibfield  {author} {\bibinfo {author} {\bibfnamefont {M.}~\bibnamefont
  {Lavelle}}\ and\ \bibinfo {author} {\bibfnamefont {D.}~\bibnamefont
  {McMullan}},\ }\href {\doibase 10.1016/j.physletb.2014.11.014} {\bibfield
  {journal} {\bibinfo  {journal} {Phys. Lett.}\ }\textbf {\bibinfo {volume}
  {B739}},\ \bibinfo {pages} {421} (\bibinfo {year} {2014})},\ \Eprint
  {http://arxiv.org/abs/1407.1279} {arXiv:1407.1279 [hep-ph]} \BibitemShut
  {NoStop}%
\bibitem [{\citenamefont {Lavelle}\ and\ \citenamefont
  {McMullan}(2015)}]{Lavelle:2015jxa}%
  \BibitemOpen
  \bibfield  {author} {\bibinfo {author} {\bibfnamefont {M.}~\bibnamefont
  {Lavelle}}\ and\ \bibinfo {author} {\bibfnamefont {D.}~\bibnamefont
  {McMullan}},\ }\href {\doibase 10.1103/PhysRevD.91.105022} {\bibfield
  {journal} {\bibinfo  {journal} {Phys. Rev.}\ }\textbf {\bibinfo {volume}
  {D91}},\ \bibinfo {pages} {105022} (\bibinfo {year} {2015})},\ \Eprint
  {http://arxiv.org/abs/1502.06529} {arXiv:1502.06529 [hep-ph]} \BibitemShut
  {NoStop}%
\bibitem [{\citenamefont {Fedotov}(2009)}]{Fedotov2009}%
  \BibitemOpen
  \bibfield  {author} {\bibinfo {author} {\bibfnamefont {A.~M.}\ \bibnamefont
  {Fedotov}},\ }\href {\doibase 10.1134/S1054660X09020108} {\bibfield
  {journal} {\bibinfo  {journal} {Laser Physics}\ }\textbf {\bibinfo {volume}
  {19}},\ \bibinfo {pages} {214} (\bibinfo {year} {2009})}\BibitemShut
  {NoStop}%
\bibitem [{\citenamefont {Di~Piazza}(2014)}]{DiPiazza:2013vra}%
  \BibitemOpen
  \bibfield  {author} {\bibinfo {author} {\bibfnamefont {A.}~\bibnamefont
  {Di~Piazza}},\ }\href {\doibase 10.1103/PhysRevLett.113.040402} {\bibfield
  {journal} {\bibinfo  {journal} {Phys. Rev. Lett.}\ }\textbf {\bibinfo
  {volume} {113}},\ \bibinfo {pages} {040402} (\bibinfo {year} {2014})},\
  \Eprint {http://arxiv.org/abs/1310.7856} {arXiv:1310.7856 [hep-ph]}
  \BibitemShut {NoStop}%
\bibitem [{\citenamefont {Di~Piazza}(2015)}]{DiPiazza:2015xva}%
  \BibitemOpen
  \bibfield  {author} {\bibinfo {author} {\bibfnamefont {A.}~\bibnamefont
  {Di~Piazza}},\ }\href {\doibase 10.1103/PhysRevA.91.042118} {\bibfield
  {journal} {\bibinfo  {journal} {Phys. Rev.}\ }\textbf {\bibinfo {volume}
  {A91}},\ \bibinfo {pages} {042118} (\bibinfo {year} {2015})},\ \Eprint
  {http://arxiv.org/abs/1501.06475} {arXiv:1501.06475 [hep-ph]} \BibitemShut
  {NoStop}%
\bibitem [{\citenamefont {Di~Piazza}(2017)}]{DiPiazza:2016tdf}%
  \BibitemOpen
  \bibfield  {author} {\bibinfo {author} {\bibfnamefont {A.}~\bibnamefont
  {Di~Piazza}},\ }\href {\doibase 10.1103/PhysRevA.95.032121} {\bibfield
  {journal} {\bibinfo  {journal} {Phys. Rev.}\ }\textbf {\bibinfo {volume}
  {A95}},\ \bibinfo {pages} {032121} (\bibinfo {year} {2017})},\ \Eprint
  {http://arxiv.org/abs/1612.04132} {arXiv:1612.04132 [hep-ph]} \BibitemShut
  {NoStop}%
\bibitem [{\citenamefont {Heinzl}\ and\ \citenamefont
  {Ilderton}(2017)}]{Heinzl:2017zsr}%
  \BibitemOpen
  \bibfield  {author} {\bibinfo {author} {\bibfnamefont {T.}~\bibnamefont
  {Heinzl}}\ and\ \bibinfo {author} {\bibfnamefont {A.}~\bibnamefont
  {Ilderton}},\ }\href {\doibase 10.1103/PhysRevLett.118.113202} {\bibfield
  {journal} {\bibinfo  {journal} {Phys. Rev. Lett.}\ }\textbf {\bibinfo
  {volume} {118}},\ \bibinfo {pages} {113202} (\bibinfo {year} {2017})},\
  \Eprint {http://arxiv.org/abs/1701.09166} {arXiv:1701.09166 [hep-ph]}
  \BibitemShut {NoStop}%
\bibitem [{\citenamefont {Waters}\ and\ \citenamefont
  {King}(2017)}]{Waters:2017tgl}%
  \BibitemOpen
  \bibfield  {author} {\bibinfo {author} {\bibfnamefont {W.~J.}\ \bibnamefont
  {Waters}}\ and\ \bibinfo {author} {\bibfnamefont {B.}~\bibnamefont {King}},\
  }\href@noop {} {\  (\bibinfo {year} {2017})},\ \Eprint
  {http://arxiv.org/abs/1705.08554} {arXiv:1705.08554 [physics.optics]}
  \BibitemShut {NoStop}%
\bibitem [{\citenamefont {Karbstein}\ and\ \citenamefont
  {Mosman}(2017)}]{Karbstein:2017jgh}%
  \BibitemOpen
  \bibfield  {author} {\bibinfo {author} {\bibfnamefont {F.}~\bibnamefont
  {Karbstein}}\ and\ \bibinfo {author} {\bibfnamefont {E.~A.}\ \bibnamefont
  {Mosman}},\ }\href {\doibase 10.1103/PhysRevD.96.116004} {\bibfield
  {journal} {\bibinfo  {journal} {Phys. Rev.}\ }\textbf {\bibinfo {volume}
  {D96}},\ \bibinfo {pages} {116004} (\bibinfo {year} {2017})},\ \Eprint
  {http://arxiv.org/abs/1711.06151} {arXiv:1711.06151 [hep-ph]} \BibitemShut
  {NoStop}%
\bibitem [{\citenamefont {King}\ and\ \citenamefont
  {Heinzl}(2016)}]{king2016measuring}%
  \BibitemOpen
  \bibfield  {author} {\bibinfo {author} {\bibfnamefont {B.}~\bibnamefont
  {King}}\ and\ \bibinfo {author} {\bibfnamefont {T.}~\bibnamefont {Heinzl}},\
  }\href@noop {} {\bibfield  {journal} {\bibinfo  {journal} {High Power Laser
  Science and Engineering}\ }\textbf {\bibinfo {volume} {4}} (\bibinfo {year}
  {2016})}\BibitemShut {NoStop}%
\bibitem [{\citenamefont {King}\ and\ \citenamefont
  {Elkina}(2016)}]{king2016vacuum}%
  \BibitemOpen
  \bibfield  {author} {\bibinfo {author} {\bibfnamefont {B.}~\bibnamefont
  {King}}\ and\ \bibinfo {author} {\bibfnamefont {N.}~\bibnamefont {Elkina}},\
  }\href@noop {} {\bibfield  {journal} {\bibinfo  {journal} {Physical Review
  A}\ }\textbf {\bibinfo {volume} {94}},\ \bibinfo {pages} {062102} (\bibinfo
  {year} {2016})}\BibitemShut {NoStop}%
\bibitem [{\citenamefont {Berestetskii}\ \emph {et~al.}(2012)\citenamefont
  {Berestetskii}, \citenamefont {Pitaevskii},\ and\ \citenamefont
  {Lifshitz}}]{berestetskii2012quantum}%
  \BibitemOpen
  \bibfield  {author} {\bibinfo {author} {\bibfnamefont {V.}~\bibnamefont
  {Berestetskii}}, \bibinfo {author} {\bibfnamefont {L.}~\bibnamefont
  {Pitaevskii}}, \ and\ \bibinfo {author} {\bibfnamefont {E.}~\bibnamefont
  {Lifshitz}},\ }\href {https://books.google.co.uk/books?id=Tpk-lqyr3GoC}
  {\emph {\bibinfo {title} {Quantum Electrodynamics}}},\ \bibinfo {number} {v.
  4}\ (\bibinfo  {publisher} {Elsevier Science},\ \bibinfo {year}
  {2012})\BibitemShut {NoStop}%
\bibitem [{\citenamefont {McMaster}(1954)}]{mcmasters1954pol}%
  \BibitemOpen
  \bibfield  {author} {\bibinfo {author} {\bibfnamefont {W.~H.}\ \bibnamefont
  {McMaster}},\ }\href {\doibase 10.1119/1.1933744} {\bibfield  {journal}
  {\bibinfo  {journal} {American Journal of Physics}\ }\textbf {\bibinfo
  {volume} {22}},\ \bibinfo {pages} {351} (\bibinfo {year} {1954})}\BibitemShut 
  {NoStop}%
\bibitem [{\citenamefont {Collett}(2005)}]{collett2005field}%
  \BibitemOpen
  \bibfield  {author} {\bibinfo {author} {\bibfnamefont {E.}~\bibnamefont
  {Collett}},\ }\href {https://books.google.co.uk/books?id=5lJwcCsLbLsC} {\emph
  {\bibinfo {title} {Field Guide to Polarization}}},\ Field Guide Series\
  (\bibinfo  {publisher} {Society of Photo Optical},\ \bibinfo {year}
  {2005})\BibitemShut {NoStop}%
\bibitem [{\citenamefont {Ritus}(1972)}]{Ritus:1972ky}%
  \BibitemOpen
  \bibfield  {author} {\bibinfo {author} {\bibfnamefont {V.~I.}\ \bibnamefont
  {Ritus}},\ }\href {\doibase 10.1016/0003-4916(72)90191-1} {\bibfield
  {journal} {\bibinfo  {journal} {Annals Phys.}\ }\textbf {\bibinfo {volume}
  {69}},\ \bibinfo {pages} {555} (\bibinfo {year} {1972})}\BibitemShut
  {NoStop}%
\bibitem [{\citenamefont {Eberly}\ and\ \citenamefont
  {Reiss}(1966)}]{Eberly:1966b}%
  \BibitemOpen
  \bibfield  {author} {\bibinfo {author} {\bibfnamefont {J.~H.}\ \bibnamefont
  {Eberly}}\ and\ \bibinfo {author} {\bibfnamefont {H.~R.}\ \bibnamefont
  {Reiss}},\ }\href {\doibase 10.1103/PhysRev.145.1035} {\bibfield  {journal}
  {\bibinfo  {journal} {Phys. Rev.}\ }\textbf {\bibinfo {volume} {145}},\
  \bibinfo {pages} {1035} (\bibinfo {year} {1966})}\BibitemShut {NoStop}%
\bibitem [{\citenamefont {Krainov}\ \emph {et~al.}()\citenamefont {Krainov},
  \citenamefont {Reiss},\ and\ \citenamefont {Smirnov}}]{Krainov:1997}%
  \BibitemOpen
  \bibfield  {author} {\bibinfo {author} {\bibfnamefont {V.~P.}\ \bibnamefont
  {Krainov}}, \bibinfo {author} {\bibfnamefont {H.~R.}\ \bibnamefont {Reiss}},
  \ and\ \bibinfo {author} {\bibfnamefont {B.~M.}\ \bibnamefont {Smirnov}},\
  }\href@noop {} {\ }\bibinfo {note} {{Radiative Processes in Atomic Physics,
  Wiley (1997)}}\BibitemShut {NoStop}%
\end{thebibliography}

%

\end{document}